\magnification=1100
\overfullrule=0pt
\baselineskip=20pt
\parskip=0pt
\def\dag{\dagger}
\def\del{\partial}

\def\a{\alpha}     
\def\b{\beta}      
\def\g{\gamma}     
\def\d{\delta}     
\def\e{\epsilon}   
\def\z{\zeta}      
\def\j{\eta}       
\def\q{\theta}

\def\l{\lambda}    
\def\m{\mu}	   
\def\n{\nu}        
\def\x{\xi}        
          
\def\p{\pi}        \def\P{\Pi}
\def\r{\rho}       
\def\s{\sigma}

\def\f{\phi}       
       
\def\y{\psi}       
\def\w{\omega}     \def\W{\mit\Omega}
\def\br{\langle}
\def\ke{\rangle}
\def\ve{\vert}
\def\rhoh{\hat{\rho }}

\def\ba{\cal A}
\def\bb{b}

\def\zbar{\bar{z}}
{\settabs 5 \columns
%\+&&&&CCNY-HEP-93/1\cr}
\bigskip
\centerline{\bf Bulk and Edge excitations in a $\nu =1$ quantum Hall ferromagnet }
%\centerline{\bf }
\bigskip\bigskip
\centerline{Rashmi Ray }
\bigskip

\centerline{ Physics Department, City College of the City University of New
York}
\centerline{ New York, NY 10031}
\bigskip
\centerline{\bf Abstract}
\bigskip
In this article, we shall focus on the collective
dynamics of the fermions in a $\nu = 1$ quantum Hall droplet.
Specifically, we propose to look at the quantum Hall ferromagnet.
In this system, the electron spins are ordered in the ground state
due to the exchange part of the Coulomb interaction and the Pauli exclusion
principle. The low energy excitations are ferromagnetic magnons. 
To provide a means for describing these magnons,
we shall discuss a method of introducing collective coordinates
in the Hilbert space of many-fermion systems. These collective coordinates
are bosonic in nature. They map a part of the fermionic Hilbert space into
a bosonic Hilbert space.
Using this technique, we shall interpret the magnons as bosonic collective excitations in
the Hilbert space of the many-electron Hall system.
By considering a Hall droplet of finite extent, we shall also obtain
the effective Lagrangian governing the spin collective excitations at the
edge of the sample.
\vfill
\noindent{$^1$E-mail address: ray@aps.org.}

%\noindent{$^2$E-mail address: sakita@sci.ccny.cuny.edu.}
\vfill\eject
\centerline{\bf I. Introduction}
\bigskip
Quantum Hall ferromagnets [1] have generated a considerable amount of interest
in recent years.
The fact that Hall states exhibit ferromagnetism is somewhat 
surprising. In conventional Hall systems, the electron spins are
aligned by the external magnetic field. Thus in the absence of
spontaneous alignment, the Hall states do not qualify as 
ferromagnetic states. However, particular samples, where the coupling
of the electronic spins to the external magnetic field has been rendered
negligible for a variety of reasons, do exhibit a spontaneous alignment
of the spins. These are the Hall ferromagnets.

Traditionally, the distinction between the integer and the fractional 
Hall effects has been made based on the origin of the gap in the single 
particle spectrum. In the integer effects, the gap, (which is 
approximately equal to the cyclotron gap for $g\sim 2$) is produced
by the Zeeman coupling of the electron spins to the external magnetic
field. On the other hand, in the fractional effects, the gap is produced
by the inter-electronic interactions.

There is, however, a scenario, where despite integer filling, the gap in 
the single particle spectrum is due to inter-electron interactions. 
For instance, in GaAs, the effective mass in the conduction band, which
appears in the expression for the cyclotron gap, is much smaller than
the actual mass of the electron, which appears in the expression for
the Zeeman gap, through the Bohr magneton $\mu_{B}$. Thus the cyclotron
gap increases effectively by a factor of $\sim 14$. Spin-orbit 
scattering reduces the effective $g$ factor by a factor of $\sim 5$.
Hence the Zeeman energy is vanishingly small compared to the cyclotron
energy (a factor of $\sim 70$ smaller). 

Imagine that the $g$ factor in the sample can, starting from the value
of 2, be gradually brought down to zero. Initially, the $\nu =1$ state
described a state with a uniform density of ${{B}\over{2\p }}$ electrons
per unit area, with each electron  in the spin ``up" state, say.
With the reduction of the Zeeman gap, one would expect this state to be 
rendered unstable as the two spin orientations become degenerate.
Experimentally, however, a gap is observed, indicating the $\nu =1$
state remains stable.

A rather heuristic argument may be adduced to explain this phenomenon.
In the lowest Landau level, the kinetic energy of the electrons has a fixed
value. Thus, to minimise the Coulomb energy, the spatial part of the
many-electron wave function should be totally antisymmetric. This in turn
will require, by the Pauli principle, the spin part of the wave function
to be completely symmetric. Thus, the Coulomb interaction causes the
spins to align and is instrumental in producing the observed gap. 
The state is genuinely ferromagnetic as the spins are spontaneously
aligned. More importantly, the distinction between the integer and
the fractional effects becomes somewhat blurred in this situation.

Since the ground state exhibits the spontaneous breaking of the
original global spin $SU(2)$ symmetry (for $g=0$) down to a $U(1)$
symmetry, the excitation spectrum must contain gapless Goldstone
bosons, namely, the ferromagnetic magnons.The effective Lagrangian
governing the dynamics of the magnons has been obtained previously
in a variety of ways [1],[7].

It is well known that the excitations located at the edges of finite
Hall samples play a crucial role in the physics of these samples.
The so-called edge states have been studied in considerable depth for
the integer and the fractional effects [2],[3]. In contrast, despite some
seminal work [4] on the edge excitatitons of ferromagnet Hall
samples, a systematic derivation of the effective Lagrangian 
governing these excitations seems to be in order.

In this article, we propose to use a recently developed technique [5]
for introducing bosonic collective variables as coordinates in 
many-fermion Hilbert spaces, in studying the bulk and the edge excitations
of the $\nu =1$ Hall ferromagnet.

\bigskip
\centerline{\bf I. Bosonic Effective Lagrangian for Collective
Excitations of Fermions} 
\bigskip
In this section, we shall outline a method [5] of introducing bosonic
collective coordinates in the Hilbert space of many-fermion systems.
These collective coordinates map a subspace of the full Hilbert space
into a bosonic Hilbert space. We shall further obtain, starting from
the second-quantised fermionic Lagrangian, an effective Lagrangian
governing the bosonic collective coordinates.

Let us consider a system of $N$ fermions, free to reside at 
$K$ sites, with $K \geq N$. The number of orthogonal Fock states
describing this system is obviously given by 
$$
D_{fermion} = {{K !}\over{{(K-N)!}{N!}}}\eqno (1.1).
$$
An arbitrary $N$-fermion determinantal wave function can be viewed as a
tensor with $N$ antisymmetric indices. The number of such independent
tensors is ${{K !}\over{{(K-N)!}{N!}}}$. We immediately realise that 
${{K !}\over{{(K-N)!}{N!}}}$ independent antisymmetric tensors form 
a basis of a completely antisymmetric irreducible representation of
$SU(K)$. Thus, starting from any given state in the $N$-particle 
Fock space, we may arrive at any other by $SU(K)$ rotations.

Consider the fermionic second-quantised operators
$\hat{c}_{\alpha }, \hat{c}^{\dag }_{\alpha }$ satisfying
$\{\hat{c}_{\alpha },\hat{c}^{\dag }_{\beta }=\delta_{\alpha \beta }$
where $\alpha , \beta = 1,2,\cdots K$. Consider $U \in SU(K)$ and
let $\hat{d} \equiv U\hat{c}$. Obviously, $\hat{d}$ satisfies the same
canonical anticommutation relation as $\hat{c}$. Let us now construct a state
$$
\vert U \ke \equiv \hat{d}^{\dag }_{1}\ \hat{d}^{\dag }_{2}\ \cdots 
\hat{d}^{\dag }_{N} \vert 0 \ke \eqno (1.2).
$$
As argued above, by varying $U$, we may recover all the possible
states in the Fock space. This enables us to keep the fermionic degrees
of freedom ``frozen" in a state with a fixed $U$ and to recover the
dynamics through a variation of the bosonic degree of freedom $U$.

Let us define
$$
P \equiv \int dU\ \vert U \ke \br U \vert  \eqno (1.3)
$$
where $dU$ is the appropriate Haar measure for $SU(K)$.
It can be shown that $P^{2}=P$. Thus $P$ is the projector onto the
subspace of $N$-particle determinantal wave functions. In the sequel,
we shall assume {\bf without proof} that this is the appropriate subspace
for the Hall ferromagnet.

The partition function of the system, projected onto this subspace, is
readily obtained to be
$$
Z=\int {\cal D}U\ e^{i\int dt\ \br U(t)\vert i\del_{t}-H \vert U(t) \ke }
\eqno (1.4),
$$
where $H$ is the second-quantised fermionic Hamiltonian.

Let us consider a fermionic Hamiltonian of the form
$$
H=\hat{c}^{\dag }_{\alpha }h_{\alpha \beta }\hat{c}_{\beta }\ + \ 
\hat{c}^{\dag }_{\alpha_{1}}\hat{c}^{\dag }_{\alpha_{2}}V^{\beta_{1} \beta_{2}}
_{\alpha_{1} \alpha_{2}}\hat{c}_{\beta_{2}}\hat{c}_{\beta_{1}}\eqno (1.5).
$$

Now 
$$
\br U \vert \hat{d}^{\dag }_{\alpha } \hat{d}_{\beta } \vert U \ke 
\equiv \rho_{\alpha \beta } \eqno (1.6).
$$
Obviously $\rho_{\alpha \beta } = \delta_{\alpha \beta }$ if $\alpha ,\beta 
\leq N$ and is zero otherwise.

Then, from (1.4), the effective bosonic Lagrangian emerges as
$$
L_{eff}= {\rm tr}\rho U^{\dag }i \del_{t}U -{\rm tr}\rho U^{\dag }hU
-(U\rho U^{\dag })_{\beta_{1}\alpha_{1}}(U\rho U^{\dag })_{\beta_{2}\alpha_{2}}
[V^{\beta_{1} \beta_{2}}
_{\alpha_{1} \alpha_{2}}-V^{\beta_{2} \beta_{1}}
_{\alpha_{1} \alpha_{2}}] \eqno (1.7).
$$

Lagrangians similar to this have been obtained in the literature [6].
\bigskip
\centerline{\bf II. Planar Fermions in the Lowest Landau Level} 
\bigskip
Before introducing the bosonic collective coordinates, let us briefly 
recapitulate [7] the basic physics of planar fermions subjected to a strong
magnetic field orthogonal to the plane. We assume from the onset that
the Zeeman coupling of the electron spins to this magnetic field is zero,
due to the vanishingly small value of the effective g-factor.

The single particle (s.p.) Hamiltonian is the celebrated Landau Hamiltonian.
If $\vec A$ is the gauge potential (we choose the symmetric gauge for
convenience) giving rise to the external magnetic field, we have
$$
h_{0} = {{1}\over{2m}}\bigl( \vec p - \vec A \bigr)^{2} \equiv 
{1\over{2m}}{\vec \P }^{2} \eqno (2.1)
$$
where $\P^{x} = -i\del_{x} -{{B}\over2}y,\ \P^{y} = -i\del_{y} +{{B}\over2}x$
and defining $\p \equiv {1\over{\sqrt{2B}}}\bigl( \P^{x} - i \P^{y} \bigr)$
and $\p^{\dag }$ as its complex conjugate, we have,
$$
\bigl[ \p , \p^{\dag } \bigr] = 1 .\eqno (2.2)
$$
The large degeneracy (${{B}\over{2\p }}$ states per unit area) of the
s.p. spectrum is expressed through the introduction of the guiding centre
coordinates:
$$
\hat X\equiv \hat x-{1\over B}\hat \P^y ,\qquad\hat Y\equiv \hat y+{1\over B}
\hat \P^x , \ \ \ \ \ \ [\hat X,\hat Y]={i\over B}\ .\eqno (2.3)
$$
The holomorphic combination and its complex conjugate,
$$
\hat a\equiv \sqrt {B\over2}(\hat X+i\hat Y) , \ \ \ \ \hat
a^\dag \equiv \sqrt {B\over2}(\hat X-i\hat Y) \ ,\eqno (2.4)
$$
satisfy
$$[\hat \p ,\hat \p^\dag ]=1,\ \ \ \
[\hat a, \hat a^\dag ]=1,\ \ \ \
[\hat a, \hat \p ]=
[\hat a, \hat \p^\dag ]=[\hat a^\dag, \hat \p ]=
[\hat a^\dag, \hat \p^\dag ]=0\ .\eqno (2.5)
$$
These are related to the coordinate operators by
$$
{\sqrt{B\over 2}}(\hat x +i\hat y )\equiv \hat z =
\hat a -i \hat\p^{\dag} ,\ \ \
{\sqrt{B\over 2}}(\hat x -i\hat y )\equiv \hat{\zbar} =
\hat{ a}^{\dag} +i \hat\p \ .\eqno (2.6)
$$
Of course $[\hat z ,\hat{\zbar }]=0$ .

The eigenbasis of $h_0$ may be taken to be $\vert n,l \ke $, where
$n=0,1,2 \cdots \infty $ and the same is true for $l$. The index
$n$ is called the Landau level index. $n=0$ corresponds to the
lowest Landau level (L.L.L.).
We have
$$
\hat \p \vert n,l \ke = \sqrt{n} \vert n-1,l \ke ,\ \hat \p^{\dag } 
\vert n,l \ke 
= \sqrt{n+1} \vert n+1,l \ke \eqno (2.7)
$$
and 
$$
\hat a \vert n,l \ke = \sqrt{l} \vert n, l-1 \ke , \ \hat a^{\dag } 
\vert n,l \ke = \sqrt{l+1} \vert n,l+1 \ke . \eqno (2.8)
$$
Thus,
$$
\hat \p^{\dag } \hat \p \vert n,l \ke = n \vert n,l \ke ,\ 
\hat a^{\dag } \hat a \vert n,l \ke = l \vert n,l \ke . \eqno (2.9)
$$
We can also define a coherent state basis for $\hat a$ [3]. If
$$
\vert \x \ke \equiv e^{\x {\hat a}^{\dag }} \vert 0 \ke \eqno (2.10)
$$
we can easily check that $\hat a \vert \x \ke = \x \vert \x \ke $.
The inner product of two coherent states is given by 
$\br \j \vert \x \ke = e^{\bar \j \x } $ and the resolution of the
identity is $\int d^2 \x e^{-\vert \x \vert^2 } \vert \x \ke 
\br \x \vert = I $, where $d^2 \x \equiv {{d\ Re \x d\ Im \x }\over{\p }}
$.
The coherent state basis is related to the $\vert l \ke $ basis through
$\br l \vert \x \ke = {{\x^{l}}\over{\sqrt{l}}}$.
The L.L.L. wave function is given by 
$$
\br \vec x \vert 0,l\ke = \sqrt{{{B}\over{2\p }}}{1\over{\sqrt l}}
e^{-{1\over {2}}\vert z \vert^2 } {\bar z}^{l} \eqno (2.11)
$$
in the $\vert l \ke $ basis and
$$
\br \vec x \vert 0,\x \ke = \sqrt{{{B}\over{2\p }}}
e^{-{1\over 2}\vert z \vert^2 + \bar z \x } \eqno (2.12)
$$
in the coherent state basis.

A smooth function of $\hat z , \hat{\zbar }$, may be expanded in the
following manner:
$$
A(\hat z , \hat{\zbar }) = A(\hat a - i\hat{\p^{\dag }},\hat{a^{\dag }}+i
\hat p)=\sum_{p,q}{{(-i^{p})}{i^{q}}\over{p!\ q!}}(\hat{\p^{\dag }})^{p}
(\hat \p )^{q} \ddag \ \del^{p}_{z}\ \del^{q}_{\bar z}A(z,\bar z)\vert_
{z=\hat a , \bar z = \hat{a^{\dag }}}\  \ddag . \eqno (2.13)
$$
Here, the symbol $\ddag \cdots \ddag $ indicates that since the 
$\hat \p , \hat{\p^{\dag }}$ have been normal ordered, the 
$\hat a, \hat{a^{\dag }}$ are automatically {\bf anti-normal ordered}.
Now when such a function is projected onto the L.L.L., only the term
with $p=0,q=0$ in the sum over $p,q$  survives. Thus, in the L.L.L.,
the function $A(\hat x,\hat y)\rightarrow \ddag \ A(\hat a, \hat{a^{\dag }})
\ \ddag$. 
It is instructive to express this anti-normal ordered operator in the coherent 
state basis. Let
$$
\ddag \ A \ \ddag \equiv \sum_{p,q} {1\over{p!\ q!}}(\hat a)^{p}\ (\hat{a^{\dag }}
)^{q}A_{pq}. \eqno (2.14)
$$
We now insert the resolution of the identity in the coherent state basis
between the $\hat a$s and the $\hat{a^{\dag }}$s in $(2.14)$.
We then obtain
$$
\ddag \ A \ \ddag = \int d^2 \x e^{-\vert \x \vert^{2}}
\vert \x \ke A(\x ,\bar{\x })\br \bar{\x }
\vert \eqno (2.15)
$$
where $A(\x ,\bar{\x })\equiv \sum_{p,q} {1\over{p!\ q!}}
A_{pq} (\x )^{p}(\bar{\x })^{q}$.
Thus the product of two individually anti-normal ordered operators is
given by
$$
\ddag A \ddag \ \ddag B \ddag = \int d^2 \x e^{-\vert \x \vert^{2}} 
\vert \x \ke A(\x ,\bar{\x })
* B(\x ,\bar{\x })\br \bar{\x }
\vert \eqno (2.16)
$$
where the $* $ product is given by
$$
A(\x ,\bar{\x })* B(\x ,\bar{\x }) \equiv \sum^{\infty }_{n=0}
{{(-1)^{n}}\over{n!}}\del^{n}_{\bar \x }A\ \del^{n}_{\x }B . \eqno (2.17)
$$
The star product is associative in that $(A* \ B)* \ C = A*\ (B* \ C)$.
This concept of the star product should be very familiar to the aficionados of
non-commutative field theories. In fact, the field theory of fermions in the
L.L.L. is an instance of such a field theory, where the non-commutativity
is restricted to the spatial coordinates.

Upto this point, we have considered only the s.p. Landau Hamiltonian.
However, there are other contributions to the many particle Hamiltonian,
which we shall now discuss.

The interaction between the electrons is the Coulomb interaction, whose
contribution, as indicated in $(1.7)$, splits naturally into a direct
part and an exchange part. It is the exchange part which is instrumental
in producing ferromagnetic behaviour in the Hall droplet.

Projected on to the L.L.L., the Coulomb term is written as:
$$
H_{c} = {1\over 2}\int d^2z_{1}\ d^{2}z_{2}\ e^{-\vert z_{1} \vert^{2}
-\vert z_{2} \vert^{2}}\ \bar{\y }_{\a }(z_{1})\bar{\y }_{\b }(z_{2})
V(\sqrt{{2\over{B}}}\vert z_{1}-z_{2} \vert )\y_{\b }(\bar{z_{2}})
\y_{\a }(\bar{z_{1}}) \eqno (2.18)
$$
where $V$ is the Coulomb interaction and $\y_{\a }$ are the second
quantised electron operators with spin $\a $, projected onto the L.L.L..
Let $\vert \y_{\a } \ke \ke $ be the abstract
notation for the second quantised electron operator, with 
$\br \vec x \vert \y_{\a } \ke \ke \equiv \y_{\a }(\vec x)$ being
the corresponding field operator. Then projection to the L.L.L.
entails:
$$
\br \vec x \vert \y_{\a } \ke \ke \equiv \y_{\a }(\vec x) 
\rightarrow \br \vec x \vert \sum^{\infty }_{l=0}\vert 0,l \ke 
\br 0,l \vert \y_{\a } \ke \ke . \eqno (2.19)
$$
We define $c_{\a }(l) \equiv \br 0,l \vert \y \ke \ke $. This is the
operator that destroys an electron with index $l$ and spin $\a $.
Then upon using $(2.11)$, we have 
$$
\y_{\a }(\vec x) \to \sqrt{{{B}\over{2\p }}}\sum^{\infty }_{l=0}
{1\over{\sqrt l}}
e^{-{1\over {2}}\vert z \vert^2 } {\bar z}^{l} c_{\a }(l) 
\equiv e^{-{1\over {2}}\vert z \vert^2 }\y_{\a }(\bar{z}).
\eqno (2.20)
$$
This is the $\y (\bar z)$ that has been used in  $(2.18)$.

Apart from the Coulomb interaction, we require that the electrons
be confined to a finite portion of the plane. This entails
the introduction of a suitable confining potential [5].

Let us introduce a radially symmetric confining potential $v(r)$ that
confines $N$ electrons in a droplet of radius $R$.
$$
v(\hat z, \hat{\zbar }) = \g \vert \hat{z} \vert^2 , \eqno (2.21)
$$
where $\g $ is the strength of the confining potential.

In view of the fact that the coordinate operators do not commute
when projected to the L.L.L.. The confining potential
is really the Hamiltonian of a one dimensional harmonic oscillator,
with the coordinate operators acting as canonically conjugate variables.
The eigenvalues of $v$ are therefore given by
$$
\e_{n} = \g (n+{1\over 2}), n=0,1,2, \cdots \infty . \eqno (2.22)
$$
If we fill the available single particle states with $N$ particles
to form a droplet with the lowest possible energy, the energy of
this droplet would be
$$
E_{tot}\equiv \sum^{N-1}_{n=0}\ \e_{n} = \g \ {{N^2}\over {2}}.
\eqno (2.23)
$$
We know that the degeneracy in the L.L.L. is given by ${{B}\over{2\p }}$
particles per unit area. If the radius of the droplet be $R$, we have
the relation
$$
N = \bigl( {{B}\over{2\p }}\bigr) \bigl( \p R^2 \bigr ) \Rightarrow 
N = {{B R^2}\over{2}} . \eqno (2.24)
$$
Now, if the magnetic field is held fixed and we increase the number of
electrons, the droplet obviously increases in size. However, from
$(2.23)$, it is clear that $E_{tot} \sim N^2$ as this is done. Thus,
from $(2.24)$, it seems that the energy increases quadratically
with the size of the sample. However, for the energy to be a
properly extensive quantity, we require that it should be directly
proportional to the area (i.e. to $N$). The way to resolve this is to
consider a
confining potential whose strength is of order one (in $N$) in the bulk
but is of order $N$ at the boundary. This form of the confining
potential shall be seen to have a crucial significance in determining
the leading order terms in the bosonic effective Lagrangian that we
shall compute further on.

The ground state of the many electron system is formed by filling
up the s.p. states of the confining potential sequentially with
electrons. Furthermore, as is well known, the ground
state of the system is ferromagnetic for large $N$. Thus the 
electrons in the ground
state all have say, spin ``up". Let us now try to relate all this to 
the formulation of the collective theory given in section I. The 
available sites ($K$ in number) correspond to the s.p. eigenstates
of the confining potential (labelled by integers). Furthermore, each
state has two values of the spin index associated with it. Thus, in
line with what we have said in section I, the bosonic collective
fields are now expressed in terms of unitary operators belonging
to the fundamental representation of $SU(2K)$, with $K \to \infty $.
Similarly, the operator $\hat \r_{0}$ of section I is given in the present
context by
$$
\hat{\r_{0}} = \sum^{N-1}_{n=0} \vert n \ke \br n \vert \ \ \W \eqno (2.25)
$$
where $\W $ is a $2\times 2$ Hermitean matrix incorporating the
information about the spin of the many body ground state.
Since the ground state of the system
is ferromagnetic, $\W = P_{+} \equiv {1\over 2}(I+\s_{3})$.
$P_{+}$ is the projector onto the spin ``up" state.

We can express the operator $\hat \r_{0}$ in the coherent state basis.
$$
\r_{0}(\vert z \vert^{2}) \equiv e^{-\vert z \vert^{2}} \br \zbar 
\vert \hat \r_{0}(\hat a, \hat a^{\dag })\vert z \ke . \eqno (2.26)
$$
For a large number of particles 
($N \to \infty )$ [5], 
$$
\r_{0}(\vert z \vert^{2}) = \theta (N-\vert z \vert^{2})\ P_{+}. 
\eqno (2.27)
$$
This immediately tells us that in order to convert this into a 
statement about the spatial extent of the system, we should require that
$\vert z \vert^{2} = {{N}\over{R^{2}}}r^{2}$, whence the theta
function would look like $\theta (N-{{N}\over{R^{2}}}r^{2})=
\theta (R^{2} - r^{2})$. This in turn tells us that  $z\sim N$. 

Let us now, as a preview of the techniques to be presented in
reasonable detail in the following section, look at a simple
calculation. We shall look at ${\rm tr}\hat{\r_{0}}\ddag \ \hat{A}\ \ddag
\ \ddag \ \hat{B} \ \ddag $, where $A,B$ are arbitrary functions of
$\hat a, \hat{a^{\dag }}$.
By introducing the coherent states, it is given by
$$
{\rm tr}\int d^2z \ e^{-\vert z \vert^{2}} 
\ \hat{\r_{0}}\ \vert z \ke A\ * \ B \br \zbar \vert = 
{\rm tr} \int d^2z \ \r_{0}(\vert z \vert^{2})\ A\ * \ B . \eqno (2.28)
$$
In the last expression the trace is over the spin indices alone.
\bigskip
\centerline{\bf III. Effective Lagrangian for the Collective Excitations
of the Quantum Hall Ferromagnet}
\bigskip
In this section, we shall present a somewhat detailed derivation of the
bosonic effective Lagrangian given in $(1.7)$, for the case of the 
quantum Hall ferromagnet. In the sequel, we shall take all the operators
to have been projected onto the L.L.L..

Since the operator $\hat U$ is unitary, we can write it as $\hat U 
\equiv e^{i \hat A}$, where $\hat A$ is a Hermitean operator. Let
us also define the function $g(z,\zbar ) \equiv e^{iA(z,\zbar )}$.
As we have noted in the previous section, $z \sim \sqrt{N}$. Thus
a derivative with respect to $z$, acting on $U(z,\zbar )$ will carry
with it a factor of $N^{-{1\over 2}}$. For a large value of $N$, we
thus have a natural parameter to expand $U$ in.
In fact [5],  
$$
U(z,\zbar ) = g(z,\zbar )\ - \ g_{,\zbar , z}\  + O({1\over{N^2}})
\eqno (3.1)
$$
where 
$g_{,\zbar , z}\ \equiv \ i\ g \ \int^{1}_{0} d \a \ e^{-i\a A}\ 
\del_{\zbar }A\ \del_{z}e^{i\a A}$.

The effective Lagrangian may be represented as 
$$
L_{eff} = {\cal A}+{\cal B}+{\cal C}+{\cal D} \eqno (3.2)
$$
where
$$
{\cal A}\equiv 
{\rm tr}\rhoh \ {\hat U}^{\dag }i\del_{t}{\hat U} \eqno (3.3)
$$
$$
{\cal B}\equiv 
-{\rm tr}\rhoh \ {\hat U}^{\dag }{\hat v_{c}}{\hat U} \eqno (3.4)
$$
$$
{\cal C}
\equiv 
{1\over{2}}\int d\vec k\ V(\vert \vec k \vert )\ {\rm tr}\ (\rhoh {\hat U}^{\dag }
e^{-i{\hat \chi }}{\hat U}\rhoh {\hat U}^{\dag }e^{i{\hat \chi }}{\hat U})
\eqno (3.5)
$$
$$
{\cal D}\equiv 
-{1\over{2}}\int d\vec k\ V(\vert \vec k \vert )\ {\rm tr}\ (\rhoh {\hat U}^{\dag }
e^{-i{\hat \chi }}{\hat U})\ {\rm tr}\ (\rhoh {\hat U}^{\dag }e^{i{\hat \chi }}{\hat U})
\eqno (3.6)
$$
where $V(\vert \vec k \vert )$ is the Fourier transform of the Coulomb potential
and $e^{i{\hat \chi }}\equiv e^{i{\bar k}a}\ e^{ika^{\dag }}$, with $k\equiv {{1}\over
{\sqrt{2B}}}(k_{x}+ik_{y})$. The third and the fourth terms in the effective Lagrangian 
represent the contributions of the exchange and the direct parts respectively of the
Coulomb term.

Using the coherent state representation and the large $N$ limit of the density, we can
show that [8] 
$$
\eqalignno{
{\cal A}&=\int d^2z\ \theta (N-\vert z \vert^2) {\rm tr}\ P_{+}\ g^{\dag }i\del_{t}g \cr 
&+{{i}\over{2}}\int d^{2}z \ \delta(N-\vert z \vert^2){\rm tr}\ P_{+}\bigl( 
g^{\dag }(\bar z \del_{\bar z}-z\del_{z})g\ g^{\dag }\del_{t}g + [g^{\dag }z\del_{z}g,
g^{\dag }\del_{t}g] \bigr). & (3.7) \cr 
}
$$
The first term is obviously a bulk term and the second, an edge term.

Similarly,
$$
\eqalignno{
{\cal B}&=\int d^2z \ \theta (N-\vert z \vert^2) [-v_{c} + {\rm tr}\ P_{+}
(\del_{z}v_{c}g^{\dag }\del_{\bar z}g - \del_{\bar z}v_{c}g^{\dag }\del_{z}g) ]\cr 
&+\int d^{2}z \ \delta(N-\vert z \vert^2)v_{c}{\rm tr}\ P_{+}g^{\dag }\del_{\bar z}g
\ g^{\dag }\del_{z}g . & (3.8) \cr 
}
$$

$$
\eqalignno{
{\cal C}= &- {{1}\over{2l}}\sqrt{{{\pi }\over{2}}}\int d^2z\ \theta (N-\vert z \vert^2)
[{\rm tr}\ P_{+}g^{\dag }\del_{\bar z}g g^{\dag }\del_{z}g - {\rm tr}\ P_{+}g^{\dag }\del_{\bar z}g
{\rm tr}\ P_{+}g^{\dag }\del_{z}g] \cr 
&+{{1}\over{4l}}\sqrt{{{\pi }\over{2}}}\int d^2z \ \delta(N-\vert z \vert^2)
{\rm tr}\ P_{+}g^{\dag }(\bar z \del_{\bar z}-z\del_{z})g . & (3.9) \cr 
}
$$

$$
\eqalignno{
{\cal D}=&{1\over{2}}\int d^2z_{1}d^2z_{2}\ V(\vert z_{1}-z_{2} \vert ) \times \cr 
&\biggl( -\theta(N-\vert z_{1} \vert^2 ){\rm tr}P_{+}[g^{\dag }\del_{\bar z_{1}}g,
g^{\dag }\del_{z_{1}}g] + \delta(N-\vert z_{1} \vert^2){\rm tr}P_{+}g^{\dag }
(\bar z_{1}\del_{\bar z_{1}} - z_{1}\del_{z_{1}})g \biggr) \cr 
&\biggl( -\theta(N-\vert z_{2} \vert^2 ){\rm tr}P_{+}[g^{\dag }\del_{\bar z_{2}}g,
g^{\dag }\del_{z_{2}}g] + \delta(N-\vert z_{2} \vert^2){\rm tr}P_{+}g^{\dag }
(\bar z_{2}\del_{\bar z_{2}} - z_{2}\del_{z_{2}})g \biggr) . & (3.10) \cr 
}
$$
A uniform background density of ${{B}\over{2\pi }}$ has been subtracted off from the direct 
contribution. The deviation, in the excited state from the uniform density of the ground state
is given by
$$
\delta \rho(\vec x) \equiv 
\biggl( -\theta(N-\vert z \vert^2 ){\rm tr}P_{+}[g^{\dag }\del_{\bar z}g,
g^{\dag }\del_{z}g] + \delta(N-\vert z \vert^2){\rm tr}P_{+}g^{\dag }
(\bar z\del_{\bar z} - z\del_{z})g \biggr) .
$$

From the above, it is clear that the contributions to the effective Lagrangian split naturally
into bulk and edge contributions.

In obtaining (3.19) and (3.10), we have used the following relations:

$$
\br \bar z_{2} \vert \hat f \vert z_{1} \ke = e^{\bar{z_{2}} z_{1}}
\int d^2w \ d^2z \ e^{-\vert z \vert^2}e^{\bar{w}(z-z_{1})-w(\bar{z}-\bar{z_{2}})}
\br \bar z \vert \hat f \vert z \ke 
$$
and
$$
\int d^2\eta f(\eta , \bar{\eta }) \int d^2w e^{\bar{w}(\eta -z)-w(\bar{\eta }-\bar{z})}
= f(z,\bar{z}).
$$  
\bigskip
\centerline{\bf IV. Scaling of the terms in the bosonic effective 
Lagrangian with $N$} 
\bigskip
In the preceding section, we have obtained the leading contributions
to the bosonic effective Lagrangian that emerge from the underlying
microscopic fermionic action. We have considered the number of 
particles in the system, $N \gg 1$ and have developed a systematic 
derivative expansion scheme, with $1/\sqrt{N}$ as the small parameter, to
identify the leading contributions. In this section, we shall study
how the various terms scale with $N$ and rewrite the various terms
as integrals over real spatial coordinates.
In equation $(3.7)$, we note that in order to write the theta function
in terms of spatial coordinates, we have to take $z = {{\sqrt{N}}\over{R}}
\j $, where $R$ is the radius of the droplet given by $R=\sqrt{{{2N}\over
{B}}}$ and $\j \equiv x+iy$.
Then we get $\q(N-\vert z \vert^{2}) = \q(R^{2}-r^{2})$. This
just means that the bulk contribution has support inside of the droplet.
The measure $d^{2}z$ then becomes ${{B}\over{2\p }}dx\ dy$. Furthermore, we
note that since ${{B}\over{2\pi }}\ \pi R^{2} = N$, $N={{BR^{2}}\over{2}}$.
Thus, we have, from $(3.7)$,
$$
{\cal A}_{bulk}
=
i{{B}\over{2\p }}\int d\vec x \ \q(R^{2}-r^{2})\ {\rm tr} 
P_{+}\ g^{\dag }\ \del_{t}g . \eqno (4.1)
$$
This term is proportional to the area of the droplet and hence is
proportional to $N$.
Again, from $(3.7)$, we have,
$$
{\cal A}_{edge} =  
{{1}\over{8\p }}\int^{2\p }_{0} d\q \ {\rm tr}P_{+}\
\bigl( \{ g^{\dag }i\del_{\q }g,g^{\dag }i\del_{t}g \} \ + \ [g^{\dag }r\del_{r}g,g^{\dag }i\del_{t}g]
\bigr)_{r=R} . \eqno (4.2)
$$
This term is of $O(1)$ in $N$ as expected as it is a boundary term
and as such should be independent of the number of particles in the
droplet.

The contributions due to the confining potential are given in
$(3.8)$.
We have argued previously that for a proper thermodynamic limit to
exist, that is for the energy of the droplet to scale as $N$, the
confining potential should be of order one (in $N$) in the bulk.
Alternatively, $\del_{z}v \sim 1/\sqrt{N}$ in the bulk. Thus, we
see that the second term in ${\cal B}$ is an order one contribution to the bulk effective
Lagrangian. Thus it is a subleading contribution (compared to O($N$)
contributions) and may be dropped. 
Again,
$$
{\cal B}_{edge}
=
-{{\w }\over{8\p }}\int^{2\p }_{0} d\q \ {\rm tr}P_{+}
\bigl(- (g^{\dag }r\del_{r}g)^{2} + [g^{\dag }r\del_{r}g,
g^{\dag }i\del_{\q }g] + (g^{\dag }i\del_{\q }g)^{2} \bigr)_{r=R}
\eqno (4.3)
$$
where $\w \equiv {{v(r=R)}\over {N}}$, and is of order one in $N$.

Let us now turn our attention to the contributions of the Coulomb term.
$$
{\cal C}_{bulk}
=
-{1\over{2l}}{1\over{\sqrt{2\p }}}
\int d\vec x \ \q(R^{2}-r^{2})\bigl[ 
{\rm tr}P_{+}\ g^{\dag }\del_{\bar{\j }}g\ g^{\dag }\del_{\j }g\ - \ 
{\rm tr}P_{+}\ g^{\dag }\del_{\bar{\j }}g\ 
{\rm tr}P_{+}\ g^{\dag }\del_{\j }g \bigr] . \eqno (4.4)
$$
Similarly, using $\bar{\j }\del_{\bar{\j }}-\j \del_{\j }
=i\del_{\q }$, where $\q $ is the plane polar angle, we get
$$
{\cal C}_{edge}
=
{1\over{8l}}{1\over{\sqrt{2\p }}}\int^{2\p }_{0}d\q 
\ {\rm tr}P_{+}\ g^{\dag }\ i\del_{\q }\ g \vert_{r=R}. \eqno (4.5)
$$
The contribution in $(4.4)$ is proportional to the area of the
droplet and is thus of order $N$, whilst that in $(4.5)$ is of
order one in $N$.

Similar arguments can be provided for the contributions from the direct
part of the Coulomb interaction.

The the leading contributions to the effective Lagrangian, governing
the collective excitations in the bulk, are given by
$$
\eqalignno 
{
&L^{(bulk)}_{eff}=
\int_{{\cal D}}d\vec x \biggl[ {{B}\over{2\p }} {\rm tr}P_{+}
g^{\dag } i\del_{t}g -{1\over{2l}}{1\over{\sqrt{2\p }}}
\bigl( {\rm tr}P_{+}g^{\dag }\del_{\bar{\j }}g\ g^{\dag }\del_{\j }g
\ - \ {\rm tr}P_{+}g^{\dag }\del_{\bar{\j }}g\ 
{\rm tr}P_{+}g^{\dag }\del_{\j }g \bigr) \biggl] \cr 
&
+{1\over{2{\p }^{2}}}\int_{{\cal D}}d\vec x_{1}\int_{{\cal D}}d\vec x_{2}\ 
V(\vert \vec x_{1}-\vec x_{2} \vert )\ 
\bigl( {\rm tr}P_{+}[g^{\dag }\del_{\bar{\j_{1}}}g,
g^{\dag }\del_{\j_{1}}g]\bigr )\ 
\bigl( {\rm tr}P_{+}[g^{\dag }\del_{\bar{\j_{2}}}g,
g^{\dag }\del_{\j_{2}}g]\bigr ) & (4.6) \cr 
}
$$
The subscript ${\cal D}$ indicates that the integral is 
over all $r < R$.

Similarly, we obtain the effective Lagrangian governing the collective
dynamics at the edge of the droplet
$$
\eqalignno
{
L^{(edge)}_{eff}=
&\int^{2\p }_{0}d\q \ \biggl[ {{1}\over{8\p }}
{\rm tr}P_{+}\
\bigl( \{ g^{\dag }i\del_{\q }g,g^{\dag }i\del_{t}g \} \ + 
[g^{\dag }r\del_{r}g,g^{\dag }i\del_{t}g ]
\bigr)_{r=R}\cr 
&-{{\w }\over{8\p }}
{\rm tr}P_{+}
\bigl(- (g^{\dag }r\del_{r}g)^{2} + [g^{\dag }r\del_{r}g,
g^{\dag }i\del_{\q }g] + (g^{\dag }i\del_{\q }g)^{2} \bigr)_{r=R}\cr 
&-{1\over{2l}}{1\over{\sqrt{2\p }}}
{\rm tr}P_{+}\ g^{\dag }\ i\del_{\q }\ g \vert_{r=R}\biggl] \cr 
&
-{1\over{8\p }}\int^{2\p }_{0}d\q_{1}d\q_{2}\ V(\vert \vec x_{1}
-\vec x_{2} \vert )\ ({\rm tr}P_{+}g^{\dag }i\del_{\q_{1}}g)
({\rm tr}P_{+}g^{\dag }i\del_{\q_{2}}g)\vert_{r_{1},r_{2}
=R} .
 & (4.7) \cr 
}
$$ 

At this point, we note that if $\hat{\r_{0}}=\sum^{\infty }_{n=0}
\vert n \ke \br n \vert \ P_{+}$, all the available s.p. states
in the L.L.L. would have been filled and the effective Lagrangian
would have been entirely a bulk effective Lagrangian 
(as the droplet would
in this case fill the entire plane). 
In fact 
$$\r_{0}(\vert z \vert^{2}) \equiv e^{-\vert z \vert^{2}}
\br \zbar \vert \hat{\r_{0}} \vert z \ke = P_{+}. \eqno (4.8)
$$
Thus, the effective Lagrangian would be given by $(4.6)$ with 
the support of the integral over $x,y$ extending over the entire
droplet. 
Namely,
$$
\eqalignno
{
&L^{(bulk)}_{eff}=
\int d\vec x \biggl[ {{B}\over{2\p }} {\rm tr}P_{+}
g^{\dag } i\del_{t}g -{1\over{2l}}{1\over{\sqrt{2\p }}}
\bigl( {\rm tr}P_{+}g^{\dag }\del_{\bar{\j }}g\ g^{\dag }\del_{\j }g
\ - \ {\rm tr}P_{+}g^{\dag }\del_{\bar{\j }}g\
{\rm tr}P_{+}g^{\dag }\del_{\j }g \bigr) \biggl] \cr
&
+{1\over{2{\p }^{2}}}\int_{{\cal D}}d\vec x_{1}\int_{{\cal D}}d\vec x_{2}\
v_{c}(\vert \vec x_{1}-\vec x_{2} \vert )\
\bigl( {\rm tr}P_{+}[g^{\dag }\del_{\bar{\j_{1}}}g,
g^{\dag }\del_{\j_{1}}g]\bigr )\
\bigl( {\rm tr}P_{+}[g^{\dag }\del_{\bar{\j_{2}}}g,
g^{\dag }\del_{\j_{2}}g]\bigr ) & (4.9) \cr
}
$$
This is the effective Lagrangian that has been discussed extensively
in the literature, in the context of ferromagnetic magnons in the 
Hall ferromagnet [1],[7].
\bigskip
\centerline{\bf V. Simplifying the effective Lagrangian}
\bigskip
In this section, we shall look closely at the various contributions to the
bulk and the edge effective Lagrangians and comment on their relative
importance. We shall look separately at the the bulk and the edge 
contributions.

In terms of the familiar Euler angles, we can parametrise $g$ as
$$
g\equiv e^{-i{{\phi }\over{2}}\sigma_{3}}\ e^{-i{{\theta }\over{2}}\sigma_{2}}\ 
e^{-i{{\chi }\over{2}}\sigma_{3}}
\eqno (5.1)
$$
with 
$$
g\sigma_{3}g^{\dag } \equiv \hat m \cdot \vec \sigma \ , \ {\hat m}^{2} = 1 \eqno (5.2)
$$
where
$$
\hat m = (\sin \theta \cos \phi , \sin \theta \sin \phi , \cos \theta ) .
$$

Again, we may parametrise $g$ in a more conventional manner as
$$
\eqalignno{
g & \equiv e^{i\vec A \cdot \vec \sigma } \cr 
g^{\dag }i\del_{\mu }g & = -\vec \sigma \cdot (\del_{\mu }\vec A + \vec A \times \del_{\mu }
\vec A ) . & (5.3) \cr 
}
$$
where we have retained upto the quadratic in $\vec A$. Furthermore, we shall assume the following
ansatz for $\vec A$,
$$
\vec A = {{1}\over {2}}(m_{2}, -m_{1}, 0) \eqno (5.4)
$$
and consider $m_{3} \simeq 1$, with $\vert \vec m_{T} \vert \ll 1$, where $\vec m_{T} \equiv
(m_{1},m_{2},0)$.

Then the various contributions to the bulk effective Lagrangian may be written as:

$$
\eqalignno{
{\cal A}_{bulk} & = -{{1}\over{2}}{{B}\over{2\pi }}\int_{D} d\vec x (1-\cos \theta )\del_{t} \phi \cr 
& = {{1}\over{2}}{{B}\over{2\pi }}\int_{D} d\vec x \int^{1}_{0} d\lambda {\hat m}_{\lambda }
(\vec x,t) \cdot [\del_{t}{\hat m}_{\lambda }(\vec x,t) \times \del_{\lambda }
{\hat m}_{\lambda }(\vec x,t)] & (5.5) \cr 
}
$$
where ${\hat m}_{\lambda } \equiv {\hat m}(\lambda \theta ,\phi )$. 
We require ${\hat m}_{\lambda =0}={\hat e}_{3}$ and ${\hat m}_{\lambda =1}={\hat m}$. Explicitly,
we may choose 
$$ {\hat m}_{\lambda }(\theta ,\phi ) = \bigl( \sin \lambda \theta \sin \phi ,
\sin \lambda \theta \cos \phi , \cos \lambda \theta \bigr) .$$ 
The second line in $(5.5)$ is the well-known
Wess-Zumino term and is proportional to the area of the droplet. The suffix $D$ indicates
that the integral has support over the entire area of the droplet.

As we have argued before, we may drop ${\cal B}_{bulk}$.

$$
{\cal C}_{bulk} = -{1\over{32l\sqrt{2\pi }}}\int_{D}d\vec x (\del_{\alpha }{\hat m})^{2}
-{{1}\over{4l}}\sqrt{{{\pi }\over{2}}}\int_{D}d\vec x \rho_{p}(\vec x,t) \eqno (5.6)
$$
where $\rho_{p} \equiv -{1\over{8\pi }}\epsilon_{\alpha \beta }{\hat m}\cdot 
(\del_{\alpha }{\hat m} \times \del_{\beta }{\hat m})$ is the well known Pontryagin index
density. The integral of $\rho_{p}$ over all space gives an integer.

$$
{\cal D}_{bulk} = -{1\over{2}}\int_{D} d\vec x_{1} d\vec x_{2} \rho_{p}(\vec x_{1},t)
V(\vert \vec x_{1} - \vec x_{2} \vert )\rho_{p}(\vec x_{2},t) \eqno (5.7).
$$

Thus, the bulk effective Lagrangian is given by:
$$
\eqalignno{
L^{(bulk)}_{eff} & = \int_{D} d\vec x \bigl[ {{B}\over{4\pi }} \int^{1}_{0}d\lambda 
{\hat m}_{\lambda }\cdot [\del_{t}{\hat m}_{\lambda }\times \del_{\lambda }{\hat m}_{\lambda }]
-{1\over{32l\sqrt{2\pi }}}(\del_{\alpha }{\hat m})^{2}-{{1}\over{4l}}\sqrt{{{\pi }\over{2}}}\rho_{p} 
\bigr] \cr 
&-{1\over{2}}\int_{D}d\vec x_{1}d\vec x_{2} \rho_{p}(\vec x_{1})V(\vert \vec x_{1} - \vec x_{2} \vert )
\rho_{p}(\vec x_{2}) . & (5.8) \cr 
}
$$

This Lagrangian has been obtained in a variety of ways in the literature. It governs the dynamics
of the ferromagnetic magnons [1],[7].

Let us now look at the effective Lagrangian governing the edge excitations. To simplify matters,
we shall focus on those excitations that satisfy the condition $\del_{r}{\hat m}(r,\theta ,t)
\vert_{r=R} = 0$.

Then,
$$
{\cal A}_{edge} \simeq {1\over{16\pi }}\int^{2\pi }_{0}d\theta [\del_{t}{\hat m} \cdot 
\del_{\theta }{\hat m} ]_{r=R} . \eqno (5.9)
$$

Further,
$$
{\cal B}_{edge} \simeq -{{\omega }\over{32 \pi }}\int^{2\pi }_{0}d\theta [(\del_{\theta }
{\hat m} )^{2} ]_{r=R} \eqno (5.10)
$$
where $\omega \equiv {{v_{c}(r=R)}\over{N}}$.

Interestingly enough, {\bf to the leading order}, the contributions of the Coulomb interaction
to the Lagrangian for the edge, are zero.

Thus,
$$
L^{(edge)}_{eff} = {1\over{16 \pi }}\int^{2\pi }_{0}d\theta [\del_{t}{\hat m} 
\cdot \del_{\theta }{\hat m} - {{\omega }\over{2}}(\del_{\theta }{\hat m} )^{2}]_{r=R} . \eqno (5.11)
$$

These edge excitations are obviously chiral in nature.

We see that even though the bulk modes (ferromagnetic magnons) and the edge modes (chiral excitations)
are both gapless, they owe their dynamics to completely different sources. For the bulk modes,
which are the Goldstone modes corresponding to a spontaneous breaking of the global spin symmetry
$SU(2) \rightarrow U(1)$, the exchange part of the Coulomb interaction is not only crucial to their
dynamics, but is truly their {\it raison d'etre}.

On the other hand, the chiral edge excitations are somewhat generic to confined Hall fluids.
Let us consider a specific instance. We shall consider a conventional $\nu =1$ Hall droplet
where the spin degrees of freedom are completely frozen by the Zeeman term. In this case,
$\rho(\vert z \vert^{2}) = \theta (N-\vert z \vert^{2})$ and $g$ is simply a $U(1)$ phase.
The edge contribution to to the effective Lagrangian can be easily shown to be
$$
L^{(edge)}_{eff}={1\over{2 \pi }}\int^{2\pi }_{0}d\theta [\del_{\theta }\Phi \del_{t}\Phi  -
{{\omega }\over{2}}(\del_{\theta }\Phi )^{2}] . \eqno (5.12)
$$
This is the familiar Lagrangian obtained in the literature [9] for a case where the Coulomb interaction
is known to be unimportant. The corresponding bulk contribution trivially vanishes.
This should convince us that the edge excitations are extremely ubiquitous and as such do not owe
their existence to the Coulomb interaction.
\bigskip
\centerline{\bf VI. Conclusions}
\bigskip
In this article, we have utilised the bosonisation technique introduced in [5] to discuss the
collective excitations of a finite-sized Hall sample where $g \rightarrow 0$. For a large sample,
(Area $ \gg {{1}\over{B}}$), we know that these excitations are [1],[7] ferromagnetic magnons
whose dynamics is governed by the exchange part of the Coulomb interaction between the electrons.
For a finite sample, we also need to consider the dynamics of the edge. 

We have shown that the effective Lagrangian for the collective excitations  splits naturally
into two pieces, one having support in the bulk of the droplet and the other at the edge.

The bulk effective Lagrangian coincides with that computed for an infinite sample [1],[7].
The edge excitations, which are chiral in nature, are to the leading order, unaffected by the
Coulomb interaction between the electrons. The mere fact that the electrons are confined to
a droplet suffices to produce chiral edge excitations.
\vfill\eject
\bigskip
\centerline{\bf Acknowledgements} 
\bigskip
This work was done collaboratively with Prof. B. Sakita. I hereby acknowledge his unstinting
help. I also thank Prof. D. Karabali for useful discussions and the organisers of the
SCES-y2k conference for inviting me to give a talk.
\bigskip
\centerline{\bf References}
\bigskip
\item{[1]} S.L. Sondhi et. al. Phys. Rev. {\bf B 47}, 16419, (1993); K. Moon et. al. Phys. Rev. 
{\bf B 51}, 5138, (1995)
\item{[2]} X.G. Wen, Phys. Rev. {\bf B 41}, 12838, (1990); M. Stone, Ann. Phys. (N.Y.), {\bf 207},
38, (1991); Phys. Rev. {\bf B 53}, 16573, (1996)
\item{[3]} R. Ray, B. Sakita, Ann. Phys. (N.Y.), {\bf 230}, 131, (1994)
\item{[4]} M.V. Milovanovic, Phys. Rev. {\bf B 57}, 9920, (1998); A. Karlhede et. al., Phys. Rev. 
{\bf B 60}, 15948, (1999); Phys. Rev. Lett. {\bf 77}, 2061, (1996); T.H. Hansson, S. Viefers, 
Phys. Rev. {\bf B 61}, 7553, (2000)
\item{[5]} B. Sakita, Phys. Lett. {\bf B 387}, 118, (1996)
\item{[6]} J.P. Blaizot, H. Orland, Phys. Rev. {\bf C 24}, 1740, (1981); H. Kuratsuji, T. Suzuki,
Prog. Theor. Phys. Suppl. {\bf 75} and {\bf 76}, 209, (1983); A. Dhar et. al., Mod. Phys. Lett. 
{\bf A 8}, 3557, (1993)
\item{[7]} R. Ray, Phys. Rev. {\bf B60}, 14154, (2000)
\item{[8]} R. Ray, B. Sakita, (in preparation)
\item{[9]} S. Iso et. al., Nucl. Phys. {\bf B 388}, 700, (1992); Phys. Lett. {\bf B 296}, 143, (1992)

\end 
\bigskip
\centerline{\bf Appendix A}
\bigskip
Let $D(U)$ denote the totally antisymmetric irreducible representation
of $SU(K)$, of dimensionality $K\choose{N}$.
Then
$$
D_{A0}\ =\ {1\over{\sqrt{N!}}}\ \e_{\b_{1} \cdots \b_{N}}\ 
U_{\a_{1}\b_{1}}\cdots U_{\a_{N}\b_{N}} \eqno (A.1)
$$
where the suffix $0$ refers to the sequence $1,1,\cdots N$ and
$A$ runs over $K\choose{N}$ values.
Further, let 
$$\vert A \ke \ \equiv \ \vert \a_{1},\a_{2}\cdots \a_{N} \ke \eqno (A.2)
$$
where $\vert \a_{1},\a_{2}\cdots \a_{N} \ke $ has been given in $(1.3)$.
Thus, from $(1.3),(1.6),(A.1)$ and $(A.2)$, we get
$$
\vert U \ke \ = \ \sum_{A} \vert A \ke D_{A0}(U) \eqno (A.3)
$$
and 
$$
\br U \vert \ = \ \sum_{A} \br A \vert D^{\ast }_{A0}(U). \eqno (A.4)
$$
We further know that
$$
\int dU\ D_{A_{1}B_{1}}(U)\ D^{\ast }_{A_{2}B_{2}}(U)\ = \ \d_{A_{1}A_{2}}\ 
\d_{B_{1}B_{2}}. \eqno (A.5)
$$
Again,
$$
\eqalignno
{
\br U_{1} \vert U_{2} \ke \ &= \ \sum_{A_{1}A_{2}}
\br A_{1} \vert A_{2} \ke \ D^{\ast }_{A_{1}0}(U_{1})\ D_{A_{2}0}(U_{2})\cr 
&= \ \sum_{A}D^{\ast }_{A0}(U_{1})\ D_{A0}(U_{2})\cr 
&= \ D_{00}(U^{-1}_{1}U_{2}) .& (A.6) \cr 
}
$$ 
Thus,
$$
\eqalignno 
{
P^{2}\ &= \ \int dU_{1}dU_{2}\ \vert U_{1} \ke \ \br U_{1}\vert 
U_{2} \ke \ \br U_{2} \vert \cr 
&= \ \int dU_{1}dU_{2}\ \vert U_{1} \ke \ D_{00}(U^{-1}_{1}U_{2})\ 
\br U_{2} \vert .& (A.7) \cr 
}
$$
Let 
$$
U_{3} \ \equiv \ U^{-1}_{1}\ U_{2} . \eqno (A.8)
$$
Then, as $dU_{1}\ = \ dU_{3}$, (Haar Measure) and $\br U_{2} \vert 
\ = \ \br U_{1}U_{3}\vert $, we get
$$
\eqalignno 
{
P^{2}\ &=\ \int dU_{1}dU_{3}\ \vert U_{1} \ke \ D_{00}(U_{3})\ 
\br U_{1}U_{3} \vert \cr 
&=\ \int dU_{1}dU_{3}\ \vert U_{1} \ke \ D_{00}(U_{3})\ 
\sum_{A} \br A \vert \ D^{\ast }_{A0}(U_{1}U_{3}) \cr 
&=\ \int dU_{1}dU_{3}\ \vert U_{1} \ke \ D_{00}(U_{3})\ 
\sum_{A_{1}A_{2}} \br A_{1} \vert \ D^{\ast }_{A_{1}A_{2}}(U_{1})\ 
D_{A_{2}0}(U_{3})\cr 
&=\ \int dU_{1}\ \vert U_{1} \ke \ \sum_{A} \br A \vert \ D^{\ast }_{A0}
(U_{1}) \cr 
&=\ P . & (A.9) \cr 
}
$$  
\bigskip
\centerline{\bf Appendix B}
\bigskip
Let
$$
\hat{\r_{0}} \ \equiv \ \sum^{N-1}_{n=0} \vert n \ke \ \br n \vert .
\eqno (B.1)
$$
Further, we define
$$
e^{-\vert z \vert^{2}}\ \br \zbar \vert \ \hat{\r_{0}} \ \vert z \ke \ 
\equiv \ \r_{0}(\vert z \vert^{2}) . \eqno (B.2)
$$
Thus,
$$
\eqalignno 
{
\r_{0}(\vert z \vert^{2})\ &= \ e^{-\vert z \vert^{2}}\ \sum^{N-1}_{n=0}
\br \zbar \vert n \ke \ \br n \vert z \ke \cr 
&=\ e^{-\vert z \vert^{2}}\ \sum^{N-1}_{n=0}{{(\vert z \vert^{2})^{n}}
\over{n!}} . & (B.3) \cr 
}
$$
Now, let
$$
F(x,N) \ \equiv \ \sum^{N-1}_{n=0}{{x^{n}}\over{n!}}\ e^{-x} . \eqno (B.4)
$$
Then,
$$
\del_{x}F(x,N) \ = \ -e^{-x}\ {{x^{N-1}}\over{(N-1)!}}\ \simeq \ 
-e^{-x}\ {{x^{N}}\over{N!}} \eqno (B.5) 
$$
for large $N$.
Using Striling's formula
$$
N!\ \approx \ \sqrt{2\p }\ e^{-N}\ (N)^{N+{1\over{2}}} \eqno (B.6)
$$
and the saddle-point method
$$
e^{-x}\ x^{N} \ = \ e^{-(x-N\ln x)} \ \simeq \ e^{-(N-N\ln N)}\ 
e^{-{{(x-N)^{2}}\over{2N}}} \eqno (B.7)
$$
we get
$$
-\del_{x}F(x,N) \ \simeq \ {1\over{\sqrt{2\p N}}}\ 
e^{-{{(x-N)^{2}}\over{2N}}} . \eqno (B.8) 
$$
Therefore,
$$
\eqalignno 
{
-\del_{\vert z \vert^{2}}\r_{0}(\vert z \vert^{2})\ &\simeq \ 
{1\over{\sqrt{2\p N}}}\ e^{-{{(\vert z \vert^{2}-N)^{2}}\over{2N}}}\cr 
&=\ {1\over{\sqrt{2\p N}}}\ e^{-{{N(\vert \s \vert^{2}-1)}\over{2}}}\cr 
&\simeq \ {1\over{N}}\ \d ({{\vert z \vert^{2}}\over{N}}-1)\cr 
&=\ \d (\vert z \vert^{2}-N)  & (B.9) \cr 
}
$$
where $\s \equiv {{\vert z \vert^{2}}\over{N}}$.
Thus,
$$
\r_{0}(\vert z \vert^{2}) \ \approx \ \q (N-\vert z \vert^{2}).
\eqno (B.10)
$$

\end

If we imagine, momentarily that the
fluctuating potentials have been switched off,
the effective Lagrangian from (2.21) yields an effective action
$$S_{\rm eff}^{(0)}=\br\y\ve \bigl(i\del_t-{E\over\sqrt{2B}}
(\hat a+{\hat a}^\dag)\bigr)\ve\y\ke\eqno(3.1)$$
where $\ve\y\ke$ is a second quantized operator.
But $(\hat a+{\hat a}^\dag)/\sqrt{2B}=\hat X$. This implies that the dominant
part of the effective Hamiltonian is
$$E\br\y\ve\hat X\ve\y\ke
=E\int_{-\infty}^{\infty}dX\ \y^\dag(X)X\y(X)\eqno (3.2)$$
where $\y(X)=\br X\ve\y\ke$ and $\{\ve X\ke\}$ is the basis of the
coordinate representation.
We define the droplet to be such that all the single particle states upto
the zero energy state are filled. So $X=0$ is the Fermi surface. But, for a
large magnetic field, $X\simeq x$, the real spatial coordinate. So, in
physical space, the edge of the droplet is at $x=0$. The ground state as
defined above is
$$\ve G\ke\equiv\prod_{X\leq0}\y^\dag(X)\ve0\ke\eqno(3.3)$$
where $\ve0\ke$ is the Fock vacuum. With respect to the ground state
we re-define the fermion operators appropriately as particle and
antiparticle (hole) operators.

Excitation of this droplet by means of the fluctuating potentials
means the destruction of an electron
within the Fermi sea and the creation of an electron outside of the Fermi
sea. In terms of the state $\ve G\ke$, this translates into the creation of
a neutral particle-antiparticle excitation from the ground state.

Given that the perturbing potential is small and slowly varying in
space-time, (this is the justification for the derivative expansion we have
performed) we would expect only those electrons within some distance
$\l\ll 1/\sqrt B$ to participate in the neutral excitations. In fact,
an expansion in $X$ about $X=0$ yields, to leading order, an action
for fermions interacting with $A$, $\bar A$, ${\ba}_0$ on the boundary of the
droplet $(X=0)$.
The neutral particle-antiparticle excitations mentioned earlier are
actually the neutral excitations around the filled Fermi sea for this
boundary action.

To extract this boundary action from $S_{\rm eff}$, we write the density
operator as
$$\eqalignno{\hat \r(z,\bar z,t)&
=\int_{-\infty}^\infty dX\int_{-\infty}^\infty
dX^\prime\ \br\y\ve X,t\ke\br X^\prime,t\ve\y\ke\br X\ve z\ke\br \bar z\ve
X^\prime\ke e^{-\ve z\ve^2}\cr
&=\int_{-\infty}^\infty dX\int_{-\infty}^\infty dX^\prime\
\y^\dag(X,t)\y(X^\prime,t)\br X\ve z\ke\br\bar z\ve
X^\prime\ke e^{-\ve z\ve^2}&(3.4)\cr}$$
where $\{\y^\dag(X,t),\y(X^\prime,t)\}=\d(X-X^\prime)$.
We note that if $X,X^\prime\leq 0$, $\y^\dag,\y$ exchange their roles with
respect to $\ve G\ke$. This implies that
$$\r(z,\bar z,t)=\int_{-\infty}^\infty dX\int_{-\infty}^\infty
dX^\prime\ [:\y^\dag(X,t)\y(X^\prime,t):+
\Theta(-X)\d(X-X^\prime)]\br X\ve z\ke\br\bar z\ve X\ke
e^{-\ve z\ve^2}\eqno(3.5)$$
where : : indicates normal ordering with respect to $\ve G\ke$. The term
independent of the fermion fields is
$$\int_{-\infty}^0dX\ \br X\ve z\ke\br\bar z\ve X\ke e^{-\ve z\ve^2}=
{1\over{\sqrt\p}}
\int_{-\infty}^{-\sqrt Bx}d\g\ \exp(-\g^2)
\sim\Theta(-x)\eqno(3.6)$$
The last expression is valid in the large B limit.
When this is inserted in (2.19),
we get an effective action involving only $\ba_\m$, called the bulk action.
$$\eqalignno{S_{\rm eff}^{\rm bulk}&\simeq
-{B\over{2\p}}\int_{-\infty}^{\infty}dt
\int_{-\infty}^{0}dx\ \int_{-\infty}^{\infty}dy\
\biggl[Ex+ {\ba}_0-{1\over{2m}}{\bb}+{E\over B}{\ba}^y
+{1\over {2B}}\e^{\m\n\r}{\ba}_\m\del_\n{\ba}_\r&(3.7)\cr
&-{{2E}\over{(2B)^2}}\biggl\{2{\ba}^x\del_y {\ba}^y+x( {\ba}^x\del_y
{\bb}-{\ba}^y\del_x{\bb})-3{\ba}^y{\bb}\biggr\}\biggr]
+{1\over{4\p}}\int_{-\infty}^{\infty}dt
\int_{-\infty}^{\infty}dy\ {\ba}^0(y,t){\ba}^y(y,t)\cr}$$
where ${\bb}\equiv\del_x{\ba}^y-\del_y{\ba}^x,$ ${\ba}^\m (y,t) \equiv {\ba}^\m
(x=0,y,t).$
We see that $S_{\rm eff}^{\rm bulk}$ contains a Chern-Simons term.
It exists only in the bulk of the droplet.
Under gauge transformation, ${\ba}^i\to {\ba}^i-\del_i\Lambda$ and
${\ba}^0\to {\ba}^0+\del_t\Lambda$,
$$\d S_{\rm eff}^{\rm bulk}=-{1\over {2\p}}\int_{-\infty}^{\infty}dt
\int_{-\infty}^{\infty}dy\ \Lambda(y,t)\bigl[\del_y{\ba}_0(y,t)+
{E\over B}\del_y{\ba}^y(y,t)\bigr]\eqno (3.8)$$
So the gauge dependence is through a boundary term.
The fermionic part of $S_{\rm eff}$ is obtained by inserting
$\colon \r(z, \bar z , t) \colon $ in place of $\r(z, \bar z , t) $
in (2.19). We expand the fermionic part of $S_{\rm eff}$ and keep only
the low momentum $(X\sim0)$ fermions. Thus we get
$$S_{\rm bdry} = \int_{-\infty}^{\infty} dt\int_{-\infty}^{\infty}
dY\ \y^\dag (Y,t) \biggl[ \{ i \del_t - {\ba}_{0}(Y,t) \} +
{E\over B} \{-i \del_{Y}-{\ba}^y(Y,t)\}\biggr] {\y} (Y,t).
\eqno(3.9)$$
Here normal ordering with respect to the ground state is implicit.
$\y(Y,t)\equiv\br Y\ve\y\ke$ is actually the Fourier transform of
$\br X\ve\y\ke$ used in (3.4), since $[\hat X,\hat Y]={i\over B}$. We
however continue to use the same symbol $\y$. Thus
$$\int_{-\infty}^{\infty} dt\ \int {{d^2 \z}} e^{-{\ve \z \ve}^2}
\y^\dag (\z, t) i\del_t {\y} (\bar \z, t) =
\int_{-\infty}^{\infty} dt\ \int_{-\infty}^{\infty} dY\ \y^\dag (Y, t)
i\del_t {\y} (Y, t).\eqno(3.10)$$
Also,
$$\br \y \ve E \hat X \ve \y \ke = i {E\over B}
\y^\dag (Y, t) \del_{Y} {\y} (Y, t).\eqno(3.11)$$
which we have used in deriving (3.9).
The neutral excitations of electron--hole pairs are the neutral
excitations of this boundary action since we have expressed the low
momentum $(X \simeq 0)$ part of the original normal ordered fermion
action as this boundary action.  The electrons in the high momentum eigen
states are unaffected as long as the momenta of the
perturbing potentials are $\ll \sqrt B$.
Since we are interested in the low energy perturbations of the droplet,
$S_{\rm eff}^{\rm bulk}$ is the net effect of the fermions inside of the
droplet. $S_{\rm bdry}$ is classically gauge invariant. But quantum
mechanically this theory is an anomalous gauge theory [7,8].
 This means that the quantized theory will not be gauge invariant. The gauge
parameter dependence of the theory can be best extracted by writing it in the
bosonic language.

By studying the bosonization of the L.L.L. fermions, the following result was
obtained in [5]. The Lagrangian is written in the form of (2.19) in terms of
the L.L.L. fermion fields, namely
$$L_{\rm eff}= \int d\vec x\  \bar \y(\vec x,t)i\del_t \y(\vec x,t) -\int d\vec
x\  \bar \y(\vec x,t) \y(\vec x,t) f(\vec x,t)\eqno(3.12)$$
where $f$ is a function of $\vec{\ba}$ and ${\ba}_0$, as given in
(2.19) and $\y(\vec x,t)$ satisfies the L.L.L. condition, $(\del_z
+{1\over2}\bar z)\y(\vec x,t)=0$.
In the bosonized form,these fields are replaced by bosonic Schr\"odinger
fields.
In the bosonized form, the Lagrangian is precisely given by (3.12), where $\y$
is
replaced by a bosonic Schr\"odinger wave field which obeys the nonlinear L.L.L.
condition,
$ \bigl(\del_z +{1\over2}\bar z -\int {{{\bar \y}\y({\vec
x}^\prime)}\over{z-z^\prime}}\bigr)\y(\vec x,t)=0$
. This L.L.L. condition can be approximately solved in the droplet
approximation. This solution is that the density ${\bar \y}\y(\vec x)$ is equal
to $B/2\p$ inside of a certain region of space (droplet) and
zero outside.
The dynamical variable is then the boundary fluctuations
of the uniform density around this ``classical'' droplet
configuration. We denote these fluctuations in the following by $\f(Y,t)$.
The first term in (3.12) gives the first term in the free chiral bosonic
Lagrangian [5],[9],[10],
in terms of $\f(Y,t)$.
Working within this droplet approximation, we obtain
$$\eqalignno{L_{\rm eff}&={{B^2}\over {8\p}}\int dYdY^\prime\ \f (Y,t)\
\e (Y-Y^\prime)\dot\f(Y^\prime,t)-{B\over {2\p}}\int dY\bigl[-{\ba}_0
-{E\over B}{\ba}^y\bigr]\f(Y)\cr
&-{{BE}\over{2\p}}\int dY\ {1\over2}\f^2(Y)-{B\over{2\p}}\int_{-\infty}^0dx
\int_{-\infty}^{\infty}dy\ \biggl[{\ba}_0-{1\over{2m}}{\bb}+Ex+{E\over B}
{\ba}^y\cr
&+{1\over {2B}}\e^{\m\n\r}{\ba}_\m\del_\n{\ba}_\r
-{{2E}\over{(2B)^2}}\biggl\{2{\ba}^x\del_y{\ba}^y+
x({\ba}^x\del_y{\bb}-{\ba}^y\del_x{\bb})-3{\ba}^y{\bb}\biggr\}
\biggr]\cr
&+{1\over{4\p}}\int_{-\infty}^{\infty}dt
\int_{-\infty}^{\infty}dy\ {\ba}^0(y,t){\ba}^y(y,t) &(3.13)\cr &}$$
where $\e(Y-Y^\prime)$ is equal to 1 for $Y>Y^\prime$ and $-1$
for $Y<Y^\prime$.
The equation of motion for $\phi(Y,t)$ is
$${{B^2}\over{4\p}}\int dY^\prime\ \e(Y-Y^\prime)\ \dot\phi (Y^\prime,t)
-{{BE}\over {2\p}}\f(Y,t)+{B\over{2\p}}({\ba}_0+{E\over B}{\ba}^y)=0
\eqno(3.14)$$
This means quantum mechanically
$$\bigl[\del_t-{E\over B}\del_Y\bigr]\bigl\langle\f(Y,t)\bigr\rangle
=-{1\over B}
\del_Y\bigr({\ba}_0+{E\over B}{\ba}^y\bigr)\eqno(3.15)$$
where $\bigl\langle\dots\bigr\rangle$ denotes the quantum mechanical
average over $\f$.
Now under gauge transformation, the change in the partition function
due to the change in the action for $\f$ is
$$\eqalignno{\bigl\langle\d S_{\rm eff}^{\rm bdry}\bigr\rangle&
-{B\over{2\p}}\int dt\int dY\ \bigl[
-\del_t\Lambda+{E\over B}\del_Y\Lambda\bigr]\bigl\langle\f(Y,t)\bigr\rangle
\cr
&=-{B\over{2\p}}\int_{-\infty}^\infty dt \int_{-\infty}^\infty dY\
\Lambda (Y,t)\bigl(\del_t-{E\over B}\del_Y\bigr)\bigl\langle\f(Y,t)\bigr
\rangle\cr
&={1\over{2\p}}\int_{-\infty}^\infty dt \int_{-\infty}^\infty dY\
\Lambda(Y,t)\bigr[\del_Y{\ba}_0+{E\over B}\del_Y{\ba}^y\bigr]&(3.16)\cr}$$
by (3.15).
So comparing (3.8) and (3.16) we see that all the
$\Lambda$-dependence precisely cancels out.
We have therefore demonstrated explicitly that all the gauge non-invariance
of the bulk, which appears as a boundary term is precisely removed by the
gauge non-invariance of the chiral bosonic action governing the surface
oscillations of the droplet.

This emphasizes the importance of the edge states in maintaining gauge
invariance in the system.
\vfill
\bigskip
\centerline{\bf V.  Conclusion}
\bigskip
In this paper we have studied the electromagnetic interactions
of a quantum Hall droplet. We integrate the higher Landau levels out
and within
the framework of a derivative-expansion scheme obtain a gauge invariant
effective action
for the electrons in the L.L.L.  We know that $2+1$ dimensional electrons in
the L.L.L.
are equivalent to $1+1$ dimensional electrons, the configuration space of
the $2+1$ dimensional system being the phase space of the $1+1$ dimensional
system, [5]. In the Thomas-Fermi picture [11], the $1+1$ dimensional electron
gas occupies a region of constant density in phase space. This region in the
$1+1$ deimensional phase space coincides precisely with the physical droplet
of electrons in the L.L.L. created by the background electrostatic
potential.
The bulk of the droplet which
corresponds to a filled fermi sea contributes an effective action, called the
bulk action, in terms
of the perturbative electromagnetic potentials.
The bulk action is not gauge invariant. This
non-invariance, however, is spurious as it is cancelled by the gauge
non-invariance
of the $1+1$ dimensional edge system. Thus the basic
mechanism for the preservation of gauge invariance as suggested in [2], [3]
and [4] is seen to be valid. Moreover, we have a well-defined and systematic
procedure for isolating the $1+1$ dimensional edge from the original $2+1$
dimensional system.

The case of the fractional Hall droplet can be similarly handled, at least
phenomenologically, by writing the bosonic density ${\bar \y}\y (\vec x) $ in
the discussion following (3.12), as $\n B/2\p$ inside of the
droplet, where $\n$ is the appropriate filling factor.
\bigskip
\centerline{\bf Acknowledgements}
\bigskip
We acknowledge useful discussions with S. Iso, D. Karabali,
P. Orland and P. Panigrahi.
This work was supported by the NSF grant PHY90-20495 and the Professional
Staff
Congress Board of Higher Education of the City
University of New York under grant
no. 6-63351.
\vfill\eject
\bigskip
\centerline{\bf References}
\bigskip
\item{[1]}  B.I. Halperin, Phys. Rev. {\bf B 25}, 2185, (1982)
\item{[2]}  X.G. Wen, Phys. Rev. {\bf B 43}, 11025, (1990) and references
therein
\item{[3]}  M. Stone, Ann. Phys. (N.Y.), {\bf 207}, 38, (1991);
Int. J. Mod. Phys. {\bf B5}, 509, (1991)
\item{[4]}  J. Fr\"ohlich \& T. Kerler, Nucl. Phys. {\bf B354}, 365, (1991)
\item{[5]}  S. Iso, D. Karabali \& B. Sakita, Nucl. Phys. {\bf B 388},
700, (1992); Phys. Lett.  {\bf B 296}, 143, (1992)
\item{[6]}  S.M. Girvin \& T. Jach, Phys. Rev. {\bf B 29}, 5617, (1983)
\item{[7]}  R. Jackiw \& R. Rajaraman, Phys. Rev. Lett. {\bf 54}, 1219, (1985)
\item{[8]}  L. Faddeev \& S. Shatashvili, Phys. Lett. {\bf 167 B} 225,(1986)
\item{[9]}  R. Floreanini \& R. Jackiw, Phys. Rev. Lett. {\bf 59},1873, (1987)
\item{[10]} J. Sonnenschein, Nucl. Phys. {\bf B309}, 752, (1988)
\item{[11]}  See e.g. C. Kittel, ``Introduction to Solid State Physics, ''
(Wiley, New York, 1986)
\end